# A novel method of speech information hiding based on 3D-Magic Matrix

Zhong-Liang Yang, Xue-Shun Peng[1]
Department of Electronic Engineering, Tsinghua University
Beijing China
e-mail: [1] chuanyepeng@163.com
[2] yfhuang@tsinghua.edu.cn

Yong-Feng Huang[2], Chin-Chen Chang
Department of Information Engineering and Computer Science, Feng Chia University
Taichung, Taiwan
e-mail: ccc@cs.ccu.edu.tw

*Abstract*—Redundant information of low-bit rate speech is extremely small, thus it's very difficult to implement large capacity steganography on the low-bit rate speech. Based on multiple vector quantization characteristics of the Line Spectrum Pair (LSP) of the speech codec, this paper proposes a steganography scheme using a 3D-Magic matrix to enlarge capacity and improve quality of speech. A cyclically moving algorithm to construct a 3D-Magic matrix for steganography is proposed in this paper, as well as an embedding and an extracting algorithm of steganography based on the 3D-Magic matrix in low-bit rate speech codec. Theoretical analysis is provided to demonstrate that the concealment and the hidden capacity are greatly improved with the proposed scheme. Experimental results show the hidden capacity is raised to 200bps in ITU-T G.723.1 codec [1]. Moreover, the quality of steganography speech in Perceptual Evaluation of Speech Quality (PESQ) reduces no more than 4%, indicating a little impact on the quality of speech. In addition, the proposed hidden scheme could prevent being detected by some steganalysis tools effectively.

## I. INTRODUCTION

Low bit-rate speech codec, for example, G723.1 is widely used in voice communication like Voice over Internet Protocol (VOIP) and mobile communications, and is becoming the main traffic of network stream. Hence, the low-bit rate speech is very suitable to be used as the hidden carrier because of three aspects of its properties: dynamic generation, interactive transmission and wide application [2-4]. In fact, at present, a lot of researchers have been conducted on steganography scheme on low bit-rate speech code. For example, Yuan [5] proposed a Quantization Index Modulation (QIM) method based on dynamic codebook, Huang [6] provided an information hidden method of direct replacement for inactive speech frame, Zhou [7] proposed a state-based steganography approach which expands the LSB method to enhance embedding capabilities, and so on. However, these methods of steganography have common shortcomings, i.e., low hidden capacity and poor concealment. Therefore, how to improve the concealment and the hidden capacity has become a big challenge for steganography in low bit-rate speech codec.

The steganography scheme based on Sudoku matrix in images have a lot of research achievements, for example, in [13] and [9], Chang proposed the idea of changing pixel information in a cover image according to both 2D-Sudoku matrix and secret information. Experimental results show that with the steganography scheme using 2D-Sudoku matrix in the cover image, the Peak Signal to Noise Ratio (PSNR) reduces to no more than 3db, and visual quality of stego-images also decreases slightly. In other words, steganography scheme using 2D-Sudoku matrix has good concealment. Moreover, the hidden capacity increases significantly. However, no research has been conducted related to applying Sudoku matrix to steganography in low bit-rate speech.

The aim of this paper is to present a steganography scheme in the low bit-rate speech which can improve the hidden capacity by using Magic matrix. By analyzing the characteristics of the low bit-rate speech coding process, we propose steganography scheme by designing a 3D-Magic matrix to embed secret information in low bit-rate speech. Thus we focus on studying the definition of the 3D-Magic matrix for steganography. In addition, we also propose an embedding algorithm and an extracting algorithm based on the 3D-Magic matrix in low bit-rate speech codec. Experimental results demonstrate that the proposed steganography scheme can achieve the aim of a hidden capacity of 200bps in the G723.1 codec.

The rest of this paper is organized as follows. In Section II, related works, such as analyzing the coding characteristics being suitable for 3D-Magic matrix of low bit-rate speech, as well as existing problems related to steganography research, are introduced. In section III, the basic framework that takes low bit-rate speech as the hidden carrier by using 3D-Magic matrix is described, and the definition of 3D-Magic matrix for steganography is suggested. Experimental results are discussed in Section IV. Finally, the conclusions and future works are drawn in Section V.

## II. RELATED WORK

By efficient compression in low bit-rate speech, there remains little redundant information. So, steganography on low bit-rate speech with high hidden capacity is still a big challenge. At present, a number of attempts have been made to study steganography on low bit-rate speech streams. Some related works are introduced below. In [10], the authors proposed a Least Significant Bit (LSB) method based on anti-noise of encoding parameters of G729.A codec. In [8] the authors suggested a least three significant bit method based on LSP parameter of G723.1 codec, this method is also based on the LSB substitution of encoded speech streams. These two methods above would lead to obvious distortion of speech, although these steganography methods can obtain relatively large hidden capacity. In [12], the authors divide the whole codebook into two groups. This QIM method is very suitable for low bit-rate speech streams including codebook vector quantization. In [11], the authors also presented a QIM-steganography scheme based on pitch period of the G723.1 codec. The advantage of QIM is that distortion of cover speech can be reduced. However, the maximum of hidden capacity for each frame speech is only 4 bits, so the hidden capacity of the QIM-steganography is too small to be used in practical applications. An improved method (CNV-QIM) was suggested in [11], which can lead to less distortion, but computational complexity of the improved method is relatively high, making it difficult to be used in practice.

To sum up, most of the present steganography schemes on low bit-rate speech codec share the principle that only one vector quantization codebook is separately selected to hide secret information. Thus those steganography schemes don't take full account of the redundancy information between the different vector quantization codebooks. Currently, none of the methods can achieve simultaneously large hidden capacity and good concealment.

Different from image and video, speech signal can be considered to be approximately stationary within 10-30ms due to the limitation of the vocal tract characteristics. By analyzing G723.1, G729.A and iLBC codec, it is found that the LSP/LSF parameters are quantitated more than once in a frame. For example, the LSP coefficients of G723.1 codec and the LSF coefficients of the iLBC codec are both quantized 3 times. Finally, by analyzing the distribution characteristic of vector quantization codebooks, it's found that distribution of LSP vector quantization codebook is monotonic and continuous within a restricted interval, which decides that the second best code word is around the best code word for an un-quantized LSP vector.

Line spectral pairs (LSP) representation was developed by Fumitada Itakura in the 1970s [15], they are used to represent linear prediction coefficients (LPC) for transmission over a channel [16]. LSPs have several properties (e.g. smaller sensitivity to quantization noise) that make them superior to direct quantization of LPCs and very useful in speech coding. Here we show a detailed analyzing description of the LSP quantitation of G723.1 codec. When the original PCM speech is encoded into the un-quantized LPC coefficients with 10 orders by the conventional Levinson-Durbin recursion, it is shown as (1). It is advantageous to convert the un-quantized LPC parameter into another type of representation called un-quantized LSP, because LSP parameters are more suitable for quantization. The un-quantized LSP vector is 10-dimension and divided into 3 sub-vectors with dimension 3, 3 and 4 respectively. Each sub-vector is quantized using an 8-bit codebook. So the size of each codebook is 256. The index of the appropriate code word in LSP codebook is searched to minimize the error criterion, which is called the best index of code word while the un-quantized LSP vector is quantized. Lots of experiments have been conducted to explore the distribution characteristic of code words in LSP codebook. Then the statistical distribution of error criterion between sub-vectors of LSP parameters and code words of the codebook in G723.1 codec are analyzed in Fig. 1.

$$A_i(z) = \frac{1}{1-\sum_{j=1}^{10} a_{ij}z^{-j}} \qquad 0 \leq i \leq 3 \qquad (1)$$

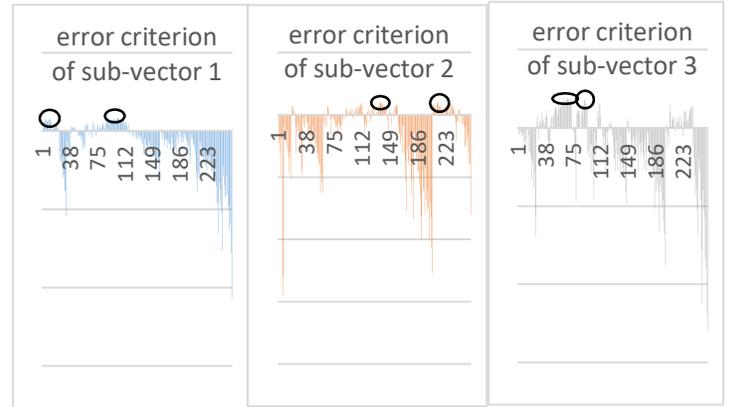

Fig. 1. Normalized distribution of error criterion between sub-vector and code words of G.723.1 LSP quantized code book

As shown in Fig. 1, an obvious characteristic can be found that the statistical distribution matches the normal distribution within a restricted interval (being marked by the black ellipse in Fig. 1, indicating that the second matched code word is nearby the best code word of codebook during LSP vector quantization. The distribution characteristic of LSP codebook is the basis of designing the steganography algorithms with Magic matrix in low bit-rate speech, because it can guarantee that the second matched code word would be searched nearby the best code word based on Magic matrix. In other words, the error caused by steganography can be achieved to be minimal.

Besides the internal characteristic of LSP vector quantization, there is another advantage for choosing 3D-Magic matrix. The 2D-Sudoku matrix has three search-pattern candidates, each candidate has distortion, and choosing a search-pattern candidate referring to a minimum distortion as the final modified result can decrease distortion [13]. Obviously, the larger the number of search-pattern candidates number, the smaller is the distortion. And 3D-Magic matrix proposed has four search-pattern candidates, so 3D-Magic matrix can reduce distortion than 2D- Sudoku matrix. This is another advantage of choosing 3D- Sudoku matrix.

## III. STEGANOGRAPHY SCHEME BASED ON SUDOKU MATRIX IN LOW BIT-RATE SPEECH CODEC

### A. General framework of steganography scheme based on 3D-Magic matrix

The proposed steganography scheme is implemented to replace the best LSP/LSF index with the second matched index based on the 3D-Magic matrix. In order to illustrate the framework of steganography clearly, here we take the G723.1 codec as an example to show how to achieve steganography scheme using 3D-Magic matrix in low bit-rate speech codec.

As shown in Fig. 2, after framing of PCM speech, LPC analysis and LSP quantization are performed, and the best LSP index numbers of 3 sub-vectors are separately searched. Thus a three-dimensional (3D) digital space of a frame speech is abstracted by 3 LSP indexes of G.723.1 codebook with 256 code words. Aimed at the 3D space, a 3D-Magic matrix is constructed for steganography. Different from 2D-Sudoku matrix used in image steganography, there are four searching patterns in order to search a matched index of code word according to the best index in 3D-Magic matrix, the similarity between code words is to compute respectively the Euclidean distance between every code word in four searching patterns and the best code word, the code word with the smallest distance is the second matched index. The second matched code word is quantized according to the secret bit "1" or "0" in steganography to decrease quantization error and improve the concealment. Moreover, the hidden capacity is also improved greatly by using a 3D-Magic matrix. But the greatest challenge of this scheme is how to define a 3D-Magic matrix for steganography on low bit-rate speech. At present, there have been not any research achievements about 3D-Magic matrix suitable for the demand of steganography.

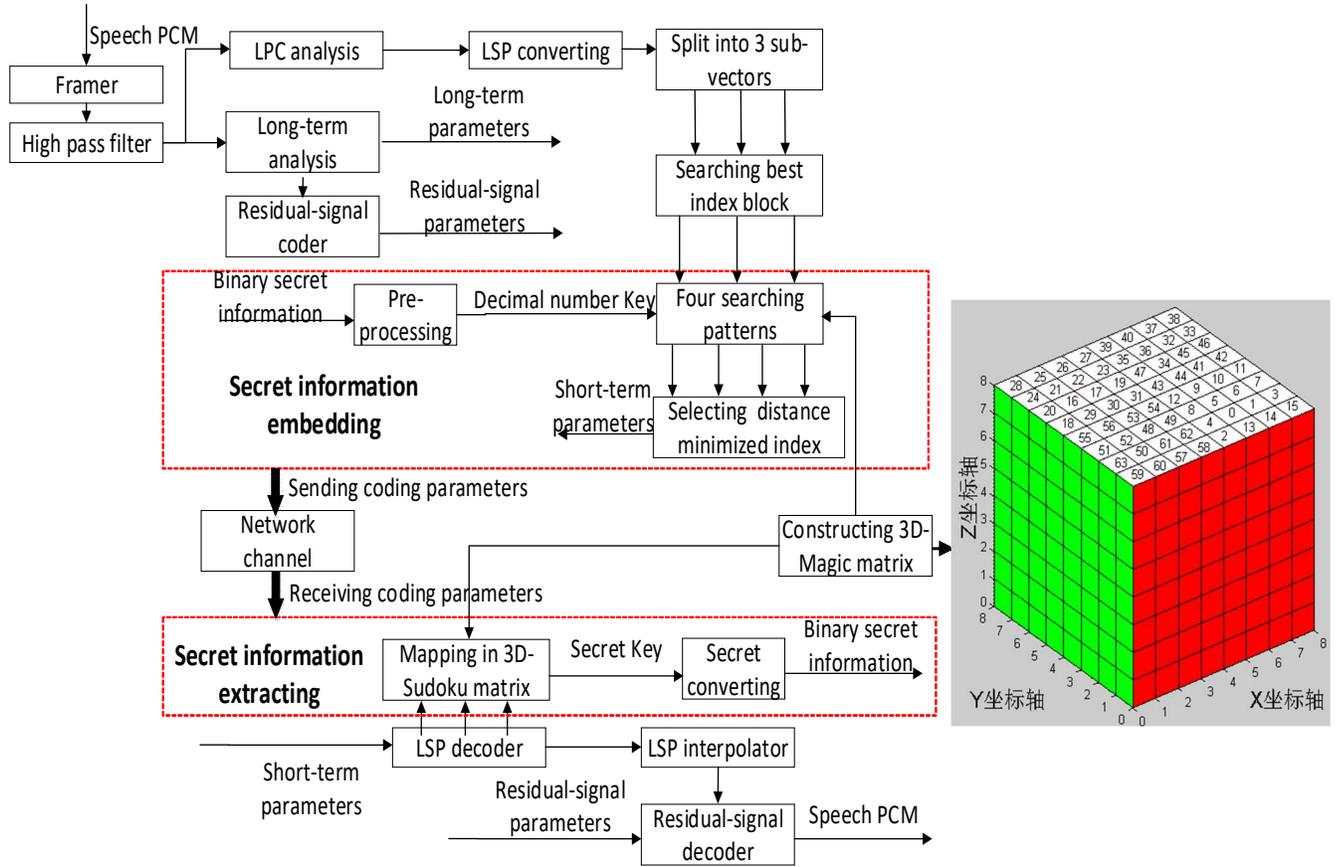

Fig. 2 The general framework of steganography scheme based on 3D-Magic matrix in G723.1 codec

### B. The definition of a 3D-Magic matrix suitable for steganography scheme

Traditional Sudoku matrix is 2-dimension, and 2D-Sudoku matrix is a 9*9 matrix which contains nine 3*3 sub-blocks for all 81 cells. Each cell is filled with the numbers 0 to 8. And some rules must also be satisfied as follows.

    (1) Each horizontal row should contain the numbers 0-8, without repeating any digitals.
    (2) Each vertical column should contain the numbers 0-8, without repeating any digitals.
    (3) Each 3*3 blocks should contain the numbers 0-8, without repeating any digitals.

Reference to the rules above, some principal rules are suggested to design 3D-Magic matrix as follows:

(1) A 3D-matrix is designed with a size of 8*8*8, which contains eight 4*4*4 sub-cubes.

(2) Each 8*8 plane (24 in total, eight Z planes, eight Y planes, eight Z planes) should contain the numbers 0-63, without repeating any digital.

(3) Each 4*4*4 cubes should contain the numbers 0-63, without repeating any digitals.

### C. Embedding algorithm based on 3D-Magic matrix in low-bit rate speech codec

The embedding algorithm in the G723.1 codec is roughly designed as follows. Firstly, the 3D-Magic matrix is randomly constructed according to Section III.B. Because the size of LSP vector codebook is 256, the 8*8*8 3D-Magic matrix is periodically expanded to 256*256*256 Sudoku matrix. Then three LSP indexes ($index_x, index_y, index_z$) are mapped into a 256*256*256 3D-Magic matrix. Finally, of the four searching patterns of secret, the mapping index with minimal quantized error is selected as the final best LSP index.

Assuming that 8*8*8 3D-Magic matrix is mapped by a special coordinate function, Secret_Value= Magic_Function(x,y,z), where x, y, z denote three coordinate values of 3D-Magic matrix, and x, y, z∈ [0, 7], Secret_Value denotes the value of cell of 3D-Magic matrix, and Secret_Value ∈ [0, 63]. The 256*256*256 Sudoku matrix coordinate function: Secret_Expand_Value=Magic_Expand (x,y,z), where x, y, z ∈ [0, 255], and Secret_Expand_Value ∈ [0, 63]. The mapping relationship of 3D-Magic matrix is shown in (3):

Secret_Expand_Value=Magic_Expand (x, y, z)=

$$\text{Magic\_Function}(x_1, y_1, z_1) \begin{cases} x_1 = x \bmod 8 \\ y_1 = y \bmod 8 \\ z_1 = z \bmod 8 \end{cases} \quad (3)$$

The embedding algorithm is described as follows.

Step 1: The long term DC component $p_{DC}$ is removed from LSP coefficients $p'$, as shown in (4), and a new DC removed LSP vector $p$ is obtained. This vector $p$ is shown in (5).

$$p = p' - p_{DC} \quad (4)$$
$$p = [p_1, p_2 \dots p_{10}] \quad (5)$$

Step 2: A first order fixed predictor, b=(12/32), is applied to the previously decoded LSP vector $\widetilde{p}$, being shown as (6), to obtain the DC removed predicted LSP vector $\bar{p}$, being shown as (7), and the residual LSP error vector $e$, the vector $e$ is calculated as (8).

$$\bar{p} = b[\widetilde{p} - p_{DC}] \quad (6)$$
$$\bar{p} = [\bar{p}_1, \bar{p}_2 \dots \bar{p}_{10}] \quad (7)$$
$$e = p - \bar{p} \quad (8)$$

Step 3: The residual LSP error vector e is divided into 3 sub-vectors with dimension 3, 3 and 4 respectively, being shown as (9).

$$e_m = [e_{1+3m}, e_{2+3m} \dots e_{K_m+3m}] \quad K_m = \begin{cases} 3, m = 0 \\ 3, m = 1 \\ 4, m = 2 \end{cases} \quad (9)$$

Step 4: Error criterion should be weighted with $W_n$. $W_n$ is a diagnal weighting matrix, determined from the unquantized LSP parameter vector $p'$ with weights defined by (10).

$$w_{j,j} = \frac{1}{\min\{p'_j - p'_{j-1}, \ p'_{j+1} - p'_j\}} \quad 2 \le j \le 9$$
$$w_{1,1} = \frac{1}{p'_2 - p'_1} \quad (10)$$
$$w_{10,10} = \frac{1}{p'_{10} - p'_9}$$

Step 5: Each m-th sub-vector is vector quantized using an 8 bit codebook, the index $l$ of the appropriate sub-vector code book entry that minimizes the error criterion $E_{l\ m}$, is selected, calculating the average of five continuous $E_{l\ m}$, as shown (11). The minimized indexes are assigned to $index_x, index_y, index_z$, $index_x \in [0, 255]$, $index_y \in [0, 255]$, and $index_z \in [0, 255]$.

$$E_{l\ m} = e_{l\ m} W_{l\ m} e_{l\ m}^T \quad \begin{cases} 0 \le m \le 2 \\ 0 \le l \le 255 \end{cases} \quad (11)$$

After obtaining three original best indexes, $index_x, index_y, index_z$, we map them into a 3D-Magic matrix, and embedding steps are shown as follows.

Step 6-1: Preprocessing of binary secret information stream, 6 bit binary secret information are converted into decimal number Key, Key∈ [0, 63].

Step 6-2: The number Key is searched in 3D matrix Magic_Expand (x,y,z) according to four searching patterns as follows. Then four different coordinates for the same number Key, ($indexi_x, indexi_y, indexi_z$), 1≤ i ≤ 4, are obtained.

① $\begin{cases} x = index_x \\ index_y - 3 \le y \le index_y + 4 \\ index_z - 3 \le z \le index_z + 4 \end{cases}$

② $\begin{cases} index_x - 3 \le x \le index_x + 4 \\ y = index_y \\ index_z - 3 \le z \le index_z + 4 \end{cases}$

③ $\begin{cases} index_x - 3 \le x \le index_x + 4 \\ index_y - 3 \le y \le index_y + 4 \\ z = index_z \end{cases}$

④ $\begin{cases} \lfloor index_x/4 \rfloor * 4 \le x \le \lfloor index_x/4 \rfloor * 4 + 3 \\ \lfloor index_y/4 \rfloor * 4 \le y \le \lfloor index_y/4 \rfloor * 4 + 3 \\ \lfloor index_z/4 \rfloor * 4 \le z \le \lfloor index_z/4 \rfloor * 4 + 3 \end{cases}$

Step 6-3: Calculating the Euclidean distance between the original coordinate and four different coordinates for the same number Key. The final best LSP, ($bestindex_x, bestindex_y, bestindex_z$) as shown in (12), is selected as the mapping relationship with the nearest Euclidean distance from the original coordinate, since the quantized error decreases as the Euclidean distance decreases.

$$(bestindex_x, bestindex_y, bestindex_z) =$$
$$\{(index_x, index_y, index_z) | \min_{i=1-4}((index_x - index_x)^2 +$$
$$(index_y - index_y)^2 + (index_z - index_z)^2)\} \quad (12)$$

In order to illustrate this embedding algorithm clearly, here we show an example. Providing the original best LSP indexes are 103,3,45, in order to display four searching patterns, these indexes should do modulo operation with 8, and become 3, 3, 5. Provided that 6 bit of binary secret information is 001111, then converted decimal number is 15. Four searching patterns are shown in Fig. 3, 4, 5, 6.

| 27 | 28 | 25 | 26 | 34 | 35 | 46 | 47 |
|---|---|---|---|---|---|---|---|
| 31 | 18 | 29 | 30 | 36 | 32 | 33 | 35 |
| 19 | 20 | 16 | 17 | 40 | 37 | 38 | 39 |
| 23 | 24 | 21 | 22 | 44 | 41 | 42 | 43 |
| 50 | 61 | 62 | 63 | 11 | 12 | 9 | 10 |
| 52 | 48 | 49 | 51 | 15 | 2 | 13 | 14 |
| 56 | 53 | 54 | 55 | 3 | 4 | 0 | 1 |
| 60 | 57 | 58 | 59 | 7 | 8 | 5 | 6 |

Fig. 3. Searching pattern plane P(5,0:7,0:7)

| 33 | 35 | 36 | 32 | 49 | 51 | 52 | 48 |
|---|---|---|---|---|---|---|---|
| 61 | 62 | 63 | 50 | 1 | 3 | 4 | 0 |
| 7 | 8 | 5 | 6 | 28 | 25 | 26 | 27 |
| 23 | 24 | 21 | 22 | 44 | 41 | 42 | 43 |
| 39 | 40 | 37 | 38 | 60 | 57 | 58 | 59 |
| 55 | 56 | 53 | 54 | 12 | 9 | 10 | 11 |
| 13 | 14 | 15 | 2 | 29 | 30 | 31 | 18 |
| 17 | 19 | 20 | 16 | 45 | 46 | 47 | 34 |

Fig. 4. Searching pattern plane P(0:7,0:7,3)

| 37 | 41 | 45 | 32 | 5 | 9 | 13 | 0 |
|---|---|---|---|---|---|---|---|
| 52 | 56 | 60 | 50 | 20 | 24 | 28 | 18 |
| 10 | 14 | 1 | 6 | 47 | 35 | 39 | 43 |
| 26 | 30 | 17 | 22 | 63 | 51 | 55 | 59 |
| 42 | 46 | 33 | 38 | 15 | 3 | 7 | 11 |
| 58 | 62 | 49 | 54 | 31 | 19 | 23 | 27 |
| 4 | 8 | 12 | 2 | 36 | 40 | 44 | 34 |
| 21 | 25 | 29 | 16 | 53 | 57 | 61 | 48 |

Fig. 5. searching pattern plane P(0:7,3,0:7)

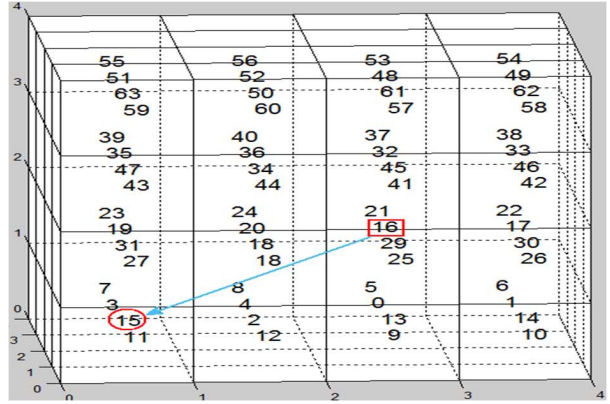

Fig. 6. Searching pattern plane P(4:7,0:3,0:3)

### D. Extracting algorithm based on 3D-Magic matrix in low-bit rate speech codec

Extracting algorithm is relatively easy. Also, the 3D-Magic matrix is constructed according to Section III.B. The three-dimensional coordinate is consist of 3 LSP indexes which are extracted through LSP decoder. Then the secret Key is obtained according to three-dimensional coordinate in the 3D-Magic matrix. Finally, the secret Key is converted into a 6-bit binary secret stream. The extracting algorithm is described as follows.

Step 1: After LSP decoder, 3 LSP indexes are extracted, denoted as $(bestindex_x, bestindex_y, bestindex_z)$.

Step 2: Constructing a 3D-Magic matrix. G723.1 decoder shares the same 3D-Magic matrix with G723.1 coder. The coordinate function is also Magic_Expand (x,y,z).

Step 3: Secret Key is obtained as three dimensional coordinate $(bestindex_x, bestindex_y, bestindex_z)$ is mapped into 3D-Magic matrix, according to (13).

$$\text{Key} = \text{Magic\_Expand}(bestindex_x, bestindex_y, bestindex_z) \quad (13)$$

In order to illustrate extracting algorithm in detail, we give an example here. Providing that the 3 LSP indexes are 103, 4, 46, the value of Magic_Expand (103, 4, 46) is 15, converted 6 bit of binary information is 001111. The secret information extracted is 001111. Then it's obvious that the extracting secret information is correct.

## IV. RESULTS AND ANALYSIS

This section reports the results and analysis. The experimental condition of steganography scheme: G723.1 codec, 6.3kbps, 30ms/frame, without silence compression, LSP quantized index, 3 sub-vectors/frames. The experimental speech data: 2000 speech sample files with PCM format as the hidden carrier for steganography. Those speech samples are classified into four groups, Chinese Man Speech (CM), Chinese Woman Speech (CW), English Man Speech (EM), and English Woman Speech (EW). Each group contains 500 pieces of speech samples, each speech sample was sampled at 8000Hz and quantized to 16 bits. To test the performance of the proposed steganography scheme, we compared the proposed steganography scheme with the schemes in paper [8] and the scheme in paper [11], and we changed a least three significant bit into a least two significant in LSP parameter in paper [8] for a fair comparison.

*A. Analysis and discussion about hidden capacity*

With the proposed steganography scheme in this paper, the changed range of LSP vector quantized index is 64 in the processing of embedding, that is, 6 ($2^6 = 64$) bit secret information can be hidden in each frame. Due to 30ms/frame, so the hidden capacity of proposed steganography scheme is 6 bit/0.03s=200bit/s. For the scheme in paper[8], the least two significant bit of LSP sub-vector quantized index is changed to embed 2-bit secret information. Due to 3 sub-vectors/frames, 3*2 bit =6 bit secret information is hidden in each frame. So the hidden capacity of paper [8] is 1 bit/0.03s=200bit/s. For the scheme in paper [11], which is essentially a parity QIM method, 1bit secret information is hidden for each sub-vector, so the hidden capacity of paper [11] is 3*1 bit/0.03s=100bit/s. In summary, when comparing the hidden capacity, the proposed steganography scheme based on 3D-Magic matrix is same as the scheme of paper [8], and is twice as big as the scheme of paper [11].

*B. Analysis and discussion about concealment*

The main evaluation criteria of concealment is the quality of stego-speech, namely imperception. Currently, there are two different evaluation methods to measure the quality of speech, i.e., objective evaluation and subjective evaluation respectively. In this study, we use Perceptual Evaluation of Speech Quality (PESQ) to evaluate the objective speech quality and use Signal to Noise Ratio (SNR) to evaluate the subjective speech quality.

PESQ was particularly developed to model subjective tests commonly used in telecommunications (e.g. ITU-T P.800) to assess the voice quality by human beings, now it becomes a worldwide applied industry standard for objective voice quality testing used by phone manufacturers, network equipment vendors and telecom operators [17]. SNR is a standard method for evaluating objective signal quality, and it's defined as the ratio of signal power to the noise power, that is

$$\text{SNR} = \frac{P_{signal}}{P_{noise}} \quad (14)$$

- Objective evaluation of speech quality

The value of PESQ is ranging from 0 to 5, and if the PESQ value is bigger, the quality of speech is better. Table I lists comparisons of changes in PESQ among the steganography scheme in paper [8], the steganography scheme in paper [11] and the proposed steganography scheme. In general, the average worsening change in PESQ of Man Speech is less than that of Woman Speech. At the same hidden capacity of 200bit/s as paper [8], the overall average worsening change in PESQ for the stego-speech files using the proposed steganography scheme is all less than the steganography scheme in paper [8]. At the condition that the hidden capacity of the proposed steganography scheme is twice as big as the scheme in paper [11], the overall average worsening change in PESQ for the stego-speech files using the proposed steganography scheme is all more than the steganography scheme of paper [11], but worsening change of both steganography schemes is less than 3%, indicating little impact on the quality of speech.

Tab.1. Comparisons of changes in PESQ among different steganography schemes

| Algorithm | | CM | CW | EM | EW | Average |
|---|---|---|---|---|---|---|
| **Standard scheme** | PESQ | 3.692 | 3.553 | 3.697 | 3.643 | - |
| | Change | - | - | - | - | - |
| **Paper[8] scheme** | PESQ | 3.447 | 3.347 | 3.481 | 3.412 | - |
| | Change | 5.82% | 5.79% | 5.84% | 6.34% | 5.85% |
| **Paper[11] scheme** | PESQ | 3.668 | 3.520 | 3.677 | 3.619 | - |
| | Change | 0.65% | 0.93% | 0.54% | 0.65% | 0.69% |
| **Proposed scheme** | PESQ | 3.586 | 3.448 | 3.599 | 3.527 | - |
| | Change | 2.87% | 2.95% | 2.65% | 3.18% | 2.91% |

- Subjective evaluation of speech quality

Fig. 7 shows comparisons of SNR among the steganography scheme in paper [8], the steganography scheme in paper [11] and the proposed steganography scheme in four groups of experimental speech data. In general, if the embedding rate is higher, the SNR value is smaller, namely, the quantized error is also bigger. According to the testing results of four sample groups, at the same hidden capacity of 200bit/s as paper [8], the SNR value for the stego-speech files using the proposed steganography scheme was larger than the steganography scheme in paper [8], indicating less impact on the quality of speech. At the condition that the hidden capacity of the proposed steganography scheme is twice as big as the scheme in paper [11], the SNR value for the stego-speech files using the proposed steganography scheme was slightly smaller than the steganography scheme in paper [11], indicating slightly more impact on the quality of speech. Of course, this could be explained that the hidden capacity is larger, the quantized error is larger.

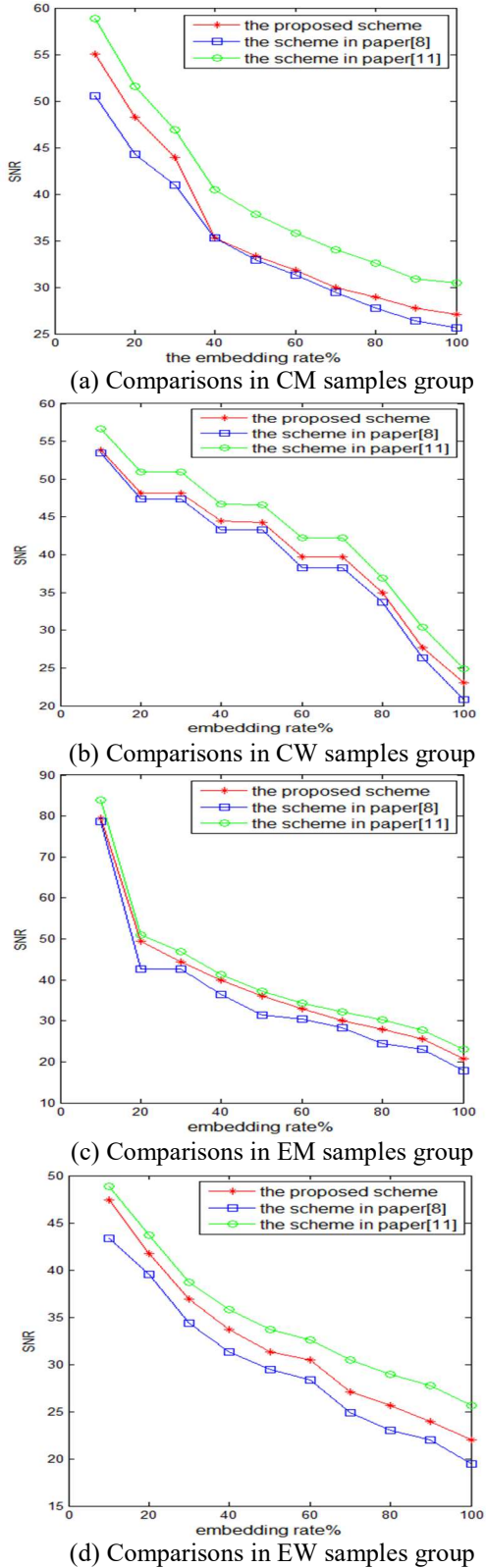

(a) Comparisons in CM samples group

(b) Comparisons in CW samples group

(c) Comparisons in EM samples group

(d) Comparisons in EW samples group

Fig. 7. SNR results with different embedding rates using different steganography schemes. (a): CM, (b): CW, (c): EM, (d): EW.

## V. CONCLUSION

This paper proposes a novel steganography scheme based on 3D-Magic matrix in low bit-rate speech codec. By taking the characteristics of 3 LSP vector quantization in low bit-rate speech codec into account, the algorithm of constructing a 3D-Magic matrix for steganography, i.e., cyclically moving algorithm, is designed. In order to reduce the noise, caused by steganography on quality of speech, we propose four searching patterns of secret and select one that quantized error is smallest among these four patterns. Moreover, we propose a method of looking up index table to speed up searching of secret. Experimental results show that the hidden capacity of using the proposed steganography scheme is 200bit/s, which is the same as the paper [8] and twice as big as the paper[11]. In addition, the worsening change in PESQ of the stego-speech by using the proposed steganography scheme was within 3%, 3% less than the paper [8], 2.2% more than paper [11]. The SNR value of our scheme was larger than the paper [8], smaller than the paper [11]. Via testing, the proposed steganography scheme has little impact on the quality of speech.

Since many different 3D-Magic matrixes for steganography can be constructed, so how to select the best matched 3D-Magic matrix is an interesting direction of the future work. Meanwhile, the detection of steganography scheme based on 3D-Magic matrix in low bit-rate speech codec is probably an open question and should be further explored.

ACKNOWLEDGMENT

This work in the paper is supported by the National Natural Science Foundation of China under the grant No. u1536113, u1405254. Fang Zhao Wu (Xueshun Peng's senior schoolfellow) provides language help during the research.